\documentstyle[a4,12pt]{article}
\def\be{\begin{equation}}
\def\ee{\end{equation}}
\def\bq{\begin{eqnarray}}
\def\eq{\end{eqnarray}}
\begin{document}
\thispagestyle{empty}
\setcounter{page}{0}
\setcounter{page}{0}
\begin{flushright}
WUE-ITP-96-012\\
MPI-PhT/96-60\\
hep-ph/9607407 
July 1996
\end{flushright}
\vspace*{\fill}
\begin{center}
{\Large\bf Calculation of the $B \rightarrow \pi $ 
Transition\\ Matrix Element in QCD}$^*$\\
\vspace{2em}
\large
A. Khodjamirian$^{a,\dagger}$ and R. R\"uckl$^{a,b}$\\
\vspace{2em}
{$^a$ \small Institut f\"ur Theoretische Physik,
Universit\"at W\"urzburg, D-97074 W\"urzburg, 
Germany}\\
{$^b$ \small Max-Planck-Institut f\"ur Physik, Werner-Heisenberg-Institut, 
D-80805 M\"unchen, Germany}\\

\end{center}
\vspace*{\fill}
 
\begin{abstract}
We describe the calculation 
of the $B\rightarrow \pi$ hadronic matrix element
from operator product 
expansion near the light-cone in QCD. In the resulting sum rules, 
the form factors $f^+$ and $F_0$ determining this matrix element
are expressed 
in terms of pion light-cone wave functions.
We present predictions for $f^+$
and $F_0$ in the region of momentum transfer 
$0 \leq p^2 < m_b^2-{\cal O}(\mbox{1 GeV}^2)$ taking into account
the wave functions up to twist 4. 
We also discuss the extrapolation of the form factors   
to higher $p^2$ and  their asymptotic 
dependence on the heavy quark mass.
\end{abstract}
 
\vspace*{\fill}
 
\begin{flushleft}

\noindent $^\dagger$ {\small on leave from 
Yerevan Physics Institute, 375036 Yerevan, Armenia } \\

\noindent $^*${\it talk presented by A. Khodjamirian at 2nd Workshop 
"Continuous Advances in QCD", TPI, University of Minnesota, 
Minneapolis, March 1996 }
\baselineskip=16pt                                 
\end{flushleft}
 
\newpage

\section{Introduction}

The matrix element of the 
$B\rightarrow \pi$ transition can be written 
in terms of two independent form factors 
$f^+$ and $f^-$:
\be
\langle\pi(q)\mid \bar{u}\gamma_\mu b\mid B(p+q)\rangle
=2f^+(p^2) q_\mu +(f^+(p^2)+f^-(p^2)) p_\mu ~,
\label{def}
\ee
where $p+q$ and $q$ denote the initial and final state four-momenta,
respectively, and $\bar{u}\gamma_\mu b$ is the relevant quark vector
current. The form factor $f^-$ is usually combined  
with $f^+$ into the scalar form factor  
\be
F_0(p^2) = f^+(p^2) +\frac{p^2}{m_B^2-m_\pi^2}f^-(p^2) ~.
\label{f0}
\ee
A reliable estimate of the amplitude 
(\ref{def}), or equivalently, of the form factors 
$f^+$ and $f^-(F_0)$ is 
very desirable, for example,  in order to extract the CKM parameter 
$V_{ub}$ from measurements of
the $B$ meson exclusive semileptonic widths 
$ B \rightarrow \pi l \nu_l$.
Obviously, such an estimate  demands 
nonperturbative methods.

The current state-of-art in calculating   
$ B \rightarrow \pi $ and other heavy-to-light form factors
makes use of lattice simulations,
heavy quark effective theory (HQET) combined with  
chiral perturbation theory, and QCD sum rules.
Each of these methods has its limitations.
For example, lattice calculations suffer from  
uncertainties connected with the necessary extrapolations
to physical quark masses. Applications of HQET 
are restricted to 
the kinematical region  of large momentum 
transfer, $ p^2 \geq m_b^2 -{\cal O}( 1 \mbox{GeV}^2)$, 
i.e. to low momentum of the pion in the $B$ rest frame. 
Also, the corrections 
to the heavy mass expansion are often substantial and cannot 
be calculated directly within HQET.

Using QCD sum rule \cite{SVZ} techniques one can approach 
the problem 
of calculating  the matrix element (\ref{def})
in two different ways.
One way is based on the operator product 
expansion (OPE) at short distances in terms of local operators. 
The second approach employs the more economical technique of 
OPE near the light-cone. 
In both methods  one has to introduce new elements
which cannot (yet) be calculated directly in QCD. 
In the more familiar short-distance approach, 
these are
vacuum condensates. In the light-cone approach, on the other hand, one 
is led to the  so-called light-cone wave functions.
These nonperturbative input quantities  are of
universal nature so that sum rules have high predictive power. 
Furthermore, QCD sum rules provide 
a unique possibility to perform calculations in the experimentally
favourable region of small 
and intermediate values of $p^2$:
\be
0 \leq p^2 < m_b^2 - {\cal O}( 1 \mbox{GeV}^2).
\label{limit}
\ee

In what follows we will report on  the results  
of a calculation of the form factors $f^+$ and $F_0$
using the light-cone technique. 
We begin in section 2 with a short  description of the method
and present some of the main results. 
In section 3, we consider the extrapolation of 
these form factors to larger $p^2$, 
beyond the limit (\ref{limit}).
Finally, in section 4 we discuss the behaviour of 
the form factors in the 
heavy mass limit.

\section{Light-cone sum rules for $ B \rightarrow \pi $ form factors }

In order to obtain the sum rules for the 
form factors $f^{\pm}$   defined in (\ref{def}),
we start from the vacuum-to-pion correlation function 
\bq
F_\mu (p,q)&=&
i \int d^4x~e^{ipx}\langle \pi(q)\mid T\{\bar{u}(x)\gamma_\mu b(x),
\bar{b}(0)i\gamma_5 d(0)\}\mid 0\rangle \nonumber\\ 
&=&F(p^2,(p+q)^2)q_\mu +\widetilde{F}(p^2,(p+q)^2)p_\mu ~.
\label{corr}
\eq
The form factors $f^+$ and $(f^+ + f^-)$
enter the dispersion relations for 
the invariant amplitudes $F$ and $\widetilde{F}$ 
through the ground state $B$--meson contribution:
\be
F(p^2,(p+q)^2)=
\frac{2f_Bm_B^2f^+(p^2)}{m_b(m_B^2-(p+q)^2)}
+\int_{s_0}^\infty ds \frac{\rho^h(p^2,s)}{s-(p+q)^2}~,
\label{hadr}
\ee
\be
\widetilde{F}(p^2,(p+q)^2)=
\frac{f_Bm_B^2 (f^+(p^2)+f^-(p^2))}{m_b(m_B^2-(p+q)^2)}
+\int_{s_0}^\infty ds \frac{\tilde{\rho}^h(p^2,s)}{s-(p+q)^2}~,
\label{hadr1}
\ee
where  the defining relation  
of the $B$ meson decay constant $f_B$, 
\be
\langle B \mid \bar{b}i\gamma_5d\mid 0\rangle=\frac{m_B^2f_B}{m_b}~,
\label{fB}
\ee
has been used. 
The integrals in (\ref{hadr}) and (\ref{hadr1})
over the spectral densities $ \rho^h$ and 
$\tilde{\rho}^h $ take into account the contributions from  
the  excited states and the continuum with $B$ meson quantum numbers, $s_0$ denoting the effective mass threshold. 

In certain momentum regions the correlation function $F_\mu$ 
can be calculated
by expanding the $T$-product of 
the currents near the light-cone $x^2 = 0 $.
Here, one uses the fact that in the region of large 
space-like momenta $(p+q)^2 <0 $ the virtual $b$ quark is far off-shell.
Simultaneously, one may  
keep the momentum $q$ at the physical 
point $q^2= m_\pi^2 $, and the momentum $p$
in the timelike region $p^2< m_b^2 -O(\mbox{1 GeV}^2)$. In
the following calculation we set $m_\pi=0$. 
The leading contribution to the OPE 
of (\ref{corr})
arises from the contraction 
of the $b$-quark operators to the free $b$-quark propagator.
Gluon emission from the virtual $b$ quark contributes at 
next-to-leading order.
After substituting the relevant expressions for the 
above,  the correlation function (\ref{corr})
is expressed in terms of vacuum-to-pion matrix elements 
of the type 
$$
\langle \pi (q)|\bar{u}(x)\Gamma_id(0)|0\rangle 
$$
\be
\langle \pi (q)|\bar{u}(x)\Gamma_j\lambda^aG^a_{\mu\nu}(vx) d(0)
|0\rangle 
\label{matrix}
\ee
where $\Gamma_{i}$ denote certain products of $\gamma$-matrices,
$G_{\mu\nu}^a$ is the gluon field tensor and $ 0\leq v \leq 1$.
The path-ordered exponential gauge factors 
of the gluon field 
are implied in (\ref{matrix}),
but omitted in correspondence to 
the Fock-Schwinger gauge which is used here. 

Each of the nonlocal operators  
in (\ref{matrix}) is equivalent to 
an infinite series of local
operators. The latter can be  
distinguished by their twist.
A detailed study of this expansion
can be found in \cite{BB}.
The concept of twist turns out to be very 
useful, because the contributions of higher twist components 
of the matrix elements (\ref{matrix}) to the correlation
function  $F_\mu$ are suppressed by powers
of the heavy quark propagator.
If the pion momentum does not vanish, which is just the case
under consideration, one has to keep 
all local operators with given twist in this series.
This would demand introducing 
an infinite number of unknown local matrix elements.
It is then better 
to work directly with the nonlocal matrix elements (\ref{matrix}).
Usually, one parametrizes the components
of different twist 
in these matrix elements 
by certain distribution functions. These distributions, known as 
the light-cone wave functions of the pion, were already introduced 
\cite{BL,CZ}
before the invention of QCD sum rules in the context of    
exclusive hard processes. They 
play a similar role  
as the vacuum condensates entering the more conventional 
short-distance sum rules. Like the condensates, they may be determined 
\cite{BF} by confronting suitable sum rules with  
experimental data.  
  
The most important matrix element 
appearing in the light-cone expansion of the correlation function 
(\ref{corr}) is 
\begin{eqnarray}
\langle\pi(q)|\bar{u}(x)\gamma_\mu\gamma_5d(0)|0\rangle =
-iq_\mu f_\pi\int_0^1du\,e^{iuqx}
\left(\varphi_\pi (u)+x^2g_1(u)+O(x^4)\right)
\nonumber \\
+
f_\pi\left( x_\mu -\frac{x^2q_\mu}{qx}\right)\int_0^1
du\,e^{iuqx}g_2(u) + ...~,
\label{phi}
\end{eqnarray}
where the r.h.s. represents the first few terms of the light-cone
expansion in $x^2$, with  
$\varphi_\pi$ being the leading twist~2 wave function, and $g_1$ 
and $g_2$ denoting twist~4 components.   
Furthermore, the matrix elements 
\be
\langle\pi(q)\mid \bar{u}(x)i\gamma_5d(0)\mid 0\rangle=
f_\pi \mu_\pi \int_0^1du~e^{iuqx}\varphi_{p}(u) + ...
\label{phip}
\ee
and 
\be
\langle\pi(q)\mid\bar{u}(x)\sigma_{\mu\nu}\gamma_5d(0)\mid 0\rangle=
i(q_\mu x_\nu -q_\nu x_\mu )\frac{f_\pi \mu_\pi}{6}
\int_0^1 du ~e^{iuqx}\varphi_{\sigma }(u) + ...\, ,
\label{phisigma}
\ee
with $\mu_\pi=m_\pi^2/(m_u + m_d)$,
involve the twist~3 wave functions $\varphi_p$ and $\varphi_\sigma$.
Contributions of twist larger than 4 are neglected. 
Referring to \cite{BKR,BBKR,KRW} for more details, we only 
mention here that the invariant amplitude $F$ 
receives the main contribution from the leading 
wave function $\varphi_\pi$, while $\varphi_\pi$ 
does not contribute to $\widetilde{F}$.
The correction to the correlation function (\ref{corr}) 
from gluon emission by the heavy quark
involves  quark-antiquark-gluon 
matrix elements of the type outlined in (\ref{matrix}).
The corresponding 3-particle wave functions are described 
in \cite{BKR,BBKR}. Direct calculation shows that the twist 3 and 4 
three-particle terms contribute to the invariant 
amplitude $F$, but not to 
the invariant amplitude $\widetilde{F}$.
Hence, to twist 4 accuracy,  the result for $\widetilde{F}$
turns out to be remarkably simple: 
\bq
\widetilde{F}_{QCD}(p^2,(p+q)^2)=f_\pi \int_0^1 \frac{du}
{m_b^2-(p+uq)^2}\left\{ 
\mu_\pi\varphi_{p}(u)
+\frac{\mu_\pi \varphi_{\sigma }(u) }{6u}
\right.
\nonumber
\\
\left. \times \Bigg[ 1-
\frac{m_b^2-p^2}{m_b^2-(p+uq)^2}\Bigg]
+ \frac{2m_b g_2(u)}{m_b^2-(p+uq)^2} \right\} ~.
\label{Ftilde}
\eq

This expression as well as the analogous result 
for $F_{QCD}$ given in \cite{BKR,BBKR}
determine the l.h.s. of 
the expressions  (\ref{hadr1}) and (\ref{hadr}) 
respectively.
In accordance with the standard procedure 
we  use quark-hadron duality and replace 
$\rho^h$ and $\tilde{\rho}^h$ at $s >s_0$ by the imaginary parts 
of $F_{QCD}$ and $\widetilde{F}_{QCD}$, respectively. 
Then, in order to 
suppress  the contributions from the excited states and from the 
continuum exponentially, 
one applies a Borel transformation in the variable $(p+q)^2$.  
The final expressions for the light-cone sum rules for 
$f^+$ and $ f^+ + f^-$ 
read
$$
f^+( p^2)=\frac{f_{\pi} m_b^2}{2f_B m_B^2}
\exp \left( \frac{m_B^2}{M^2}\right)
\Bigg\{
\int_\Delta^1\frac{du}{u} 
exp\left[-\frac{m_b^2-p^2(1-u)}{uM^2} \right]
$$
$$
\times\Bigg( \varphi_\pi(u) 
+ \frac{\mu_\pi}{m_b}
\Bigg[u\varphi_{p}(u) + \frac{
\varphi_{\sigma }(u)}{3}\left(1 + \frac{m_b^2+p^2}{2uM^2}\right) \Bigg]
- \frac{4m_b^2 g_1(u)}{ u^2M^4 }
$$
\be
+\frac{2}{uM^2} \int^u_0 g_2(v)dv
\left(1+ \frac{m_b^2+p^2}{uM^2} \right ) \Bigg)
+f^+_G(p^2,M^2) + ...,
\label{fplus}
\ee
and 
$$
f^+(p^2) + f^-(p^2)=\frac{f_\pi \mu_\pi m_b}{f_Bm_B^2 }\exp\left(
\frac{m_B^2}{M^2}\right)\Bigg\{
\int^{1}_{\Delta} 
\frac{du}{u}
\exp \left[ - \frac{m_b^2-p^2(1-u)}{u M^2}\right]
$$
\be
\times\Bigg[ 
\varphi_{p}(u) 
+ \frac{\varphi_\sigma(u)}{6u} 
\left(1- 
\frac{m_b^2-p^2}{uM^2}\right) 
+ \frac{2m_bg_2(u)}{\mu_\pi uM^2}\Bigg] +...~,
\label{fplusminus}
\ee
where $M$ is the Borel parameter and  
$\Delta = (m_b^2-p^2)/(s_0-p^2) $. In (\ref{fplus}),
$f^+_G$ denotes  the gluon corrections of twist~3 and 4 
not shown here for brevity.
The ellipses in (\ref{fplus})
and (\ref{fplusminus})
indicate threshold terms 
which emerge in addition to the integrals over $u$ after subtraction of 
the spectral density of the higher states. 
These terms are investigated in detail in \cite{KRW} 
and found to be quantitatively unimportant.

For the numerical analysis of the sum rules (\ref{fplus})
and (\ref{fplusminus}) we use 
the parameters and light-cone wave functions 
specified and discussed in detail in \cite{BKR,BBKR}. In particular, we take
$f_\pi = 132$ MeV,
$\mu_\pi= 2.02 $ GeV, $m_B=5.279$ GeV,  
$m_b=4.7$ GeV, $s_0= 35$ GeV$^2$, and $f_B=140$ MeV.
The value of $\mu_\pi$ results from the PCAC relation 
with the quark condensate density $\langle \bar{q}q \rangle= 
(-260~\mbox{MeV})^3 $,
normalized at 
the relevant scale $\mu_b =\sqrt{m_B^2-m_b^2}=2.4~\rm{GeV}$. 
As argued in \cite{BKR,BBKR}, we take the parameters $ f_B $ and $s_0$ 
from a QCD sum rule for the correlator of two 
$ \bar{b}\gamma_5 u $ currents.
For consistency, this two-point sum rule is used
without $O(\alpha_s)$ corrections, since these corrections 
are also absent in the present sum rules (\ref{fplus}) and 
(\ref{fplusminus}). The formal expressions for the pion
wave functions are collected in \cite{BBKR} and will 
not be presented again for the sake of brevity.

In order to extract reliable 
and selfconsistent values for $f^+$ and $(f^++f^-)$, 
we restrict the Borel parameter $M$ to the fiducial interval 
in which the twist~4 contributions to
(\ref{fplus}) and (\ref{fplusminus}) do not exceed 10\% , 
and simultaneously the  higher states 
do not contribute more than 30\%. 
For $0 \leq p^2< 20 $ GeV$^2$ these criteria are satisfied at  
8 GeV$^2 <M^2< $12 GeV$^2$. Within this interval both 
sum rules are found to be very stable. 

The prediction for $f^+$ obtained from 
(\ref{fplus}) at the central value,
$M^2 =10$ GeV$^2$, of the fiducial 
interval is plotted in Fig. 1. 
Combining (\ref{fplus})
and (\ref{fplusminus})
and using the definition (\ref{f0})
one easily obtains the prediction for the scalar 
form factor $F_0$. This is also shown in Fig. 1.

\section{Extrapolation to higher momentum transfer}
 
At  $p^2 \rightarrow (m_B-m_\pi)^2 $ , i.e. 
close to the kinematical threshold at zero pion recoil, the 
form factor $f^+(p^2)$ is expected to have a simple 
pole behaviour, 
\be
f^+_{pole}(p^2) = \frac{f_{B^*}g_{B^*B\pi}}{2m_{B^*}(1-p^2/m_{B^*}^2)}~,
\label{vectpole}
\ee
determined by the vector ground state $B^*$. Here, the parameters
$f_{B^*} $ and $g_{B^*B\pi}$ 
are defined by the matrix elements 
\be
\langle 0 \mid \bar{u}\gamma_\mu b\mid B^*\rangle
=m_{B^*}f_{B^*}\epsilon_\mu   ~,
\label{fB*}
\vspace{0.3cm}        
\ee
and 
\be
\langle B^{*-}(p)\pi^+(q)\mid \bar{B}^0(p+q)\rangle =
-g_{B^*B\pi}q_\mu \epsilon ^{\mu},
\label{Bstar}
\ee   
$\epsilon_\mu$ being the polarization vector of the $B^*$.
The physical argument   
for the validity of the single-pole approximation (\ref{vectpole})
relies on  the fact that the mass of the 
$B^*$ is very close to the threshold  of
the $B\rightarrow \pi$ transition.
However, there is a priori no reason 
to expect the approximation (\ref{vectpole}) to be valid
at small $p^2$, i.e. far away from this threshold.

In \cite{BBKR} we investigated this question by comparing the 
prediction of the light-cone sum rule for $f^+$ with the 
extrapolation of the single-pole approximation. The 
$B^*B\pi$ coupling was   
estimated from the correlation
function (\ref{corr}) using a double dispersion 
relation. We found the 
two descriptions  of $f^+$, (\ref{fplus})
and (\ref{vectpole}),
to be  very similar in shape and
magnitude. Hence, they smoothly match  
at
$p^2\simeq 15 \div 17~\rm{GeV}^2$ (see Fig. 1). 
From this, one may conclude
that the contributions of excited $B^*$ states to $f^+$ are
numerically not very important at  
intermediate momentum transfer.  At smaller $p^2$, 
the numerical disagreement between the pole model 
extrapolation and the light-cone sum rule prediction for $f^+$ 
increases amounting to about 50 \% at $p^2=0$. This difference  is 
insensitive to the choice of parameters, simply because 
the input used in 
the sum rule for $g_{B^*B\pi}$ is exactly the same as in the sum rule
(\ref{fplus}). 

For several reasons the form factor $F_0$ 
cannot be approximated by a single scalar pole, not even at
large $p^2$. 
Although the scalar $B$ states are still  
poorly known, it seems clear that  
the relevant distance between the lowest lying scalar pole
and the physical region of the $B\rightarrow \pi$ transition
is larger than in the case of $f^+$ and the vector $B^*$ pole. 
Therefore, excited scalar states may have substantial 
influence on $F^0$  even at  $p^2$ close to the kinematical threshold $(m_B-m_\pi)^2$. Moreover, not only the scalar but also the vector 
states in principle contribute to 
the dispersion relation for $F_0$. 

Interestingly, there exists 
a model independent 
constraint on the behaviour of the form factor $F_0$ at  
$p^2 \sim m_B^2$. 
From current algebra and PCAC one can obtain \cite{Vol} a 
Callan-Treiman type relation:  
\be
\mbox{lim}_{p^2\rightarrow m_B^2}F_0(p^2) = f_B/f_\pi~.
\label{CT}
\ee
Unfortunately, current theoretical
estimates of $f_B$ are still quite uncertain. A recent 
compilation of lattice data \cite{fb}
gives
\be
f_B/f_\pi = 1.0 \div 1.7 ~. 
\label{interval}
\ee
This is consistent with various QCD sum rule estimates 
(see e.g. \cite{fB} and \cite{fB1}),
but not yet very useful for constraining $F_0$.

\section{The heavy quark limits of $f^+$ and $(f^++f^-)$}

The light-cone sum rules presented in section 2 provide a unique 
possibility
to investigate the heavy mass dependence of the 
$B\rightarrow \pi$ transition matrix element. 
To this end, one introduces 
the scale-independent effective parameters
$\bar{\Lambda}$, $\omega_0$ and $\tau$, through the relations
$$ 
m_B = m_b+\bar{\Lambda},
$$
$$
s_0 = m_b^2 + 2m_b\omega_0 ~,
$$
\be
M^2= 2m_b\tau~,
\label{hqet}
\ee
and  uses the familiar scaling laws
for the coupling constants in the heavy quark limit:
\be
f_B = \hat{f}_B/\sqrt{m_b}, ~~~~ f_{B^*} = \hat{f}_{B^*}/\sqrt{m_b} 
\label{fBhat}
\ee

These substitutions allow to readily extract the 
leading power of the heavy mass expansion 
of (\ref{fplus}) and (\ref{fplusminus}).
We find that both sum rules have a consistent 
heavy mass expansion, that is 
the higher-twist contributions  
either have the same heavy mass behaviour as the leading twist term,
or they are suppressed by extra powers of the heavy quark mass. 
Furthermore, the heavy-mass behaviour of the form factors 
sharply differs at small and large momentum transfers.
At $p^2= 0 $ and $m_b \rightarrow \infty$ one has
\be
f^+(0)= F_0(0)\sim m_b^{-3/2}~,
\label{0limit}
\ee
\be
f^+(0) +f^-(0)\sim m_b^{-3/2}~.
\label{0limit1}
\ee
The fact that the light-cone sum rule
predicts $f^+(0) \sim m_b^{-3/2}$  
was first noticed in \cite{CZ1}. 
In contrast, at 
large momentum transfers characterized by   
\be
p^2 \sim m_b^2-2m_b\chi~,
\label{region}
\ee
where $\chi$ is finite and does not scale with $m_b$,
so that the sum rules are still valid, we get 
\bq 
f^+(p^2)
\sim m_b^{1/2}~,
\label{scal1}
\\ 
f^+(p^2) + f^-(p^2)
 \sim F_0(p^2)\sim m_b^{-1/2}~.
\label{scal2}
\eq

The sum rules thus   
nicely reproduce  
the asymptotic dependence of the form factors $f^+$
and $f^-$ on the heavy quark mass $m_b$ derived in \cite{Vol,IW}
in the 
kinematical region where the pion has small momentum in the rest  
frame of the $B$ meson. 

It should be stressed that 
there are no constraints from HQET 
when $p^2 \rightarrow 0$.
Also the pole-dominance model which is sometimes used  
for extrapolating the form factors from the region
(\ref{region}) to zero momentum transfer 
cannot be trusted outside of (\ref{region}).
As our analysis for the $f^+$ form factor  
shows, at small momentum transfer 
higher states are expected to be
important. Therefore, the change 
in the heavy mass behaviour between
small and large $p^2$ predicted by the light-cone
sum rules is very reasonable.

\section{Conclusions}

Sum rules on the light-cone are 
an economical and reliable method
for calculating exclusive hadronic amplitudes involving a single pion
(or kaon). Here,we have described    
the first complete evaluation of the $B\rightarrow \pi$ transition
matrix element in this framework.

The accuracy of the method is
conservatively estimated to be around 20-30\%. 
Currently, the main sources of uncertainty are our limited knowledge 
of the nonasymptotic
terms in the wave functions and the lack of the calculation of 
the perturbative
$\alpha_s$-corrections to the correlation function (\ref{corr}).
There is room for improvement.

In Fig. 1, we compare our predictions
for the form factors $f^+$ and $F_0$ with recent results of 
lattice calculations \cite{Lat} taken from  
\cite{Lellouch}. 
In the region of overlap, one observes 
encouraging agreement.

\section*{Acknowledgements}
A.K. is grateful to M.A. Shifman, A.I. Vainshtein, and M.B. Voloshin
for organizing a very enjoyable and fruitful Workshop
and for useful discussions. 
This work is
supported by the German Federal Ministry for Research and 
Technology (BMBF) under contract No. 05 7WZ91P(0).

\end{document}